\documentclass[aps,prb,preprint,groupedaddress]{revtex4}
\usepackage{graphicx}% Include figure files
\usepackage{dcolumn}% Align table columns on decimal point
\usepackage{bm}% bold math
\begin{document}

%\preprint{}

\title{Anisotropic photoinduced magnetism of a Rb$_{j}$Co$_{k}$[Fe(CN)$_{6}$]$_{l}$$\cdot$$n$H$_{2}$O thin film}
\author{J.-H. Park$^{a)}$}
%\author{J.-H. Park}
%\email{juhyun@phys.ufl.edu}
\author{E. \v{C}i\v{z}m\'{a}r$^{b)}$}
%\author{E. \v{C}i\v{z}m\'{a}r}
%\altaffiliation[Current address: ]{Institute of Physics, Faculty
%of Science, P. J. \v{S}af\'{a}rik University, Ko\v{s}ice,
%Slovakia.}
\author{M. W. Meisel}
\affiliation{Department of Physics and the Center for Condensed
Matter Sciences, University of Florida, Gainesville, FL
32611-8440}
\author{Y. D. Huh$^{c)}$}
%\author{Y. D. Huh$^{c)}$}
%\altaffiliation[Current address: ]{Department of Chemistry,
%Dankook University, Seoul, Korea.}
\author{F. Frye}
\author{S. Lane}
\author{D. R. Talham}
\affiliation{Department of Chemistry, University of Florida,
Gainesville, FL 32611-7200}


%\author{}
%\thanks{}
%\affiliation{}


\date{\today}

\begin{abstract}
A magneto-optically active thin film of
Rb$_{j}$Co$_{k}$[Fe(CN)$_{6}$]$_{l}$$\cdot$$n$H$_{2}$O has been
prepared using a sequential assembly method. Upon irradiation with
light and at 5~K, the net magnetization of the film increased when
the surface of the film was oriented parallel to the external
magnetic field of 0.1~T. However, when the surface of the film was
perpendicular to the field, the net magnetization \emph{decreased}
upon irradiation. The presence of dipolar fields and the
low-dimensional nature of the system are used to describe the
orientation dependence of the photoinduced magnetization. The
ability to increase or decrease the photoinduced magnetization by
changing the orientation of the system with respect to the field
is a new phenomenon that may be useful in future device
applications.\end{abstract}

%\pacs{}
%\keywords{}

\maketitle
%
%%%%%%%%%%%%%%%%%%%%%%%%%%%%%%%%%%%%%%%%%%%%%%%%%%%%%%%%%%%%%%%%%%%%%%%%%%%%%%%%%
%%%%%%%%%%%%%%%%%%%%%%%%%%%%%%%%%%%%%%%%%%%%%%%%%%%%%%%%%%%%%%%%%%%%%%%%%%%%%%%%%
The magnetic properties of thin films can be different from those
of bulk materials and may exhibit anisotropy originating from the
low-dimensional character of the systems. The high surface to
volume ratios of thin films are often advantageous in
magneto-optical studies since more photons can interact with the
system without being attenuated.\cite {Verdaguer1}  We adopted a
sequential assembly method to generate a magnetic,
optically-controllable, thin film based on a Prussian blue
analogue, a class of molecular magnets consisting of octahedral
[M(CN)$_{6}$]$^{n-}$ complexes bridged through metal ions.\cite
{Villain1}
%In particular, a film of
%Rb$_{j}$Co$_{k}$[Fe(CN)$_{6}$]$_{l}$$\cdot$$n$H$_{2}$O was
%synthesized, and
In this letter, we report the first example of a thin film in
which an increase or decrease of the photoinduced magnetization
was controlled by the orientation of the system with respect to
the applied magnetic field.

A sequential synthesis protocol \cite {Talham1, Talham2} was used
to generate a thin film whose general chemical formula was
Rb$_{j}$Co$_{k}$[Fe(CN)$_{6}$]$_{l}$$\cdot$$n$H$_{2}$O.  The
substrate, a mylar sheet \mbox{(thickness $\sim100$ $\mu$m)}, was
repeatedly immersed, for 20~cycles, in a 5 $\times$ 10$^{-3}$~M
aqueous solution of cobalt(II) nitrate and then in a mixed
solution of 2 $\times$ 10$^{-2}$~M potassium \mbox{ferricyanide}
and \mbox{1.25 $\times$ 10$^{-2}$~M} rubidium nitrate. The metal
ratios used to prepare the film yield bulk samples with formula
Rb$_{1.8}$Co$_{4}$[Fe(CN)$_{6}$]$_{3.3}$$\cdot$$13$H$_{2}$O.\cite
{Verdaguer3a} While the composition of the film has not yet been
chemically confirmed, the magneto-optical properties, which in
this series are very sensitive to composition,\cite {Verdaguer3c}
are consistent with this approximate formula. Based upon atomic
force microscope (AFM) and scanning electron microscope (SEM)
studies, the film was determined to be continuous and to have an
average thickness of about 200~nm and roughness of 50~nm. The
resulting film was cut into squares and stacked into a
polycarbonate sample holder. A bundle of optical fibers was
introduced to the stacked films and connected to a room
temperature halogen light source. Using a commercial
superconducting quantum interference device (\textsc{squid})
magnetometer, the magnetization of the sample was investigated
from 5~K to 300~K and in fields up to 7~T.

The photoinduced magnetism in Prussian blue analogues has been
extensively studied over the last decade,\cite {Verdaguer1,
Verdaguer2, Hashimoto1, Yamabe1, Hashimoto2, Verdaguer3a,
Verdaguer3b, Verdaguer3c, Verdaguer3d, Epstein1, Miller1, Abe1,
Hashimoto4, Hashimoto3, Sakata1, Park1} and the family of cobalt
hexacyanoferrate was one of the first examples in which the
diamagnetic to ferrimagnetic transition was observed upon
irradiation with visible light. More specifically, in its dark
state, the bulk compound
Rb$_{j}$Co$_{k}$[Fe(CN)$_{6}$]$_{l}$$\cdot$$n$H$_{2}$O contains a
mixture of two dominant spin configurations below
$T$$\mathrm{_{C}}$ $\sim$ 20 K, see Table 1. The relative
population of each state can be controlled by chemically tuning
the Co/Fe ratio of the sample.\cite {Verdaguer3a, Verdaguer3c}
Upon irradiation with light, an electron from the
Fe$\mathrm{^{II}}$(LS, $S = 0$) ion transfers to the
Co$\mathrm{^{III}}$(LS, $S = 0$) ion within the diamagnetic sites,
creating a ferrimagnetic spin state, namely
Fe$\mathrm{^{III}}$(LS, $S = 1/2$)--CN--Co$\mathrm{^{II}}$(HS, $S
= 3/2$), Table~1.\cite {Yamabe1, Verdaguer3d, Abe1} In other words
as a consequence of the irradiation, the net magnetization of the
compound increases, and the metastable
Fe$\mathrm{^{III}}$(LS)/Co$\mathrm{^{II}}$(HS) state is maintained
as long as the sample is kept below approximately 150~K.

The photoinduced increase of the magnetization of our film is
shown in Fig.~1(a), and this effect was observed when the surface
of the film was parallel to the external magnetic field
(H$_{\mathrm{E}}$ $\parallel$ film) of 0.1 T. This observation is
consistent with the measurements made on bulk powdered
samples.\cite {Verdaguer1, Hashimoto2, Verdaguer3a} However, under
the same experimental conditions (i.e.~the sample was field-cooled
\cite {FieldCool} to 5~K with H$_{\mathrm{E}}$ = 0.1 T, and then
subsequently exposed to light), the net magnetization
\emph{decreased} when the film was oriented perpendicular to the
field (H$_{\mathrm{E}}$ $\perp$~film), as shown in Fig.~1(c).

The anisotropy of these magneto-optical effects was sustained in
the temperature dependences of the field-cooled (fc) and
zero-field-cooled (zfc) magnetizations, see Figs.~1(b) and 1(d).
In both orientations, the divergences between the zfc and fc
magnetizations are consistent with cluster spin-glass behavior and
indicate the existence of interacting ferrimagnetic domains. As a
result of photoinduced spin creation, the temperature where the
zfc magnetization achieves its maximum value has shifted towards
higher temperatures, and this behavior can be described as an
increase in the size of the spin clusters.\cite {Epstein1, Park1}
This argument also explains the photoinduced increase of
$T$$_{\mathrm{C}}$ when H$_{\mathrm{E}}$ $\parallel$ film,
Fig.~1(b). When H$_{\mathrm{E}}$ $\perp$~film, Fig.~1(d), the
photoinduced increase of $T$$_{\mathrm{C}}$ is more subtle, as it
competes with cluster spin-glass effects.

The anisotropy of the  magneto-optical properties of our film
arises from an interplay between the low-dimensional nature of the
system and the dipolar magnetic fields generated by the
ferrimagnetic domains. This interactive relationship is shown
schematically in Fig.~2, for the case where H$_{\mathrm{E}}$
$\perp$ film.  When the sample is cooled from room temperature to
below $T$$\mathrm{_{C}}$, the primordial spin configurations
consist of ferrimagnetic domains surrounded by diamagnetic
regions, Fig.~2(a).  This arrangement arises from the random
distribution of the local chemical composition of the film.
However, structural order is maintained at least of the size of
the magnetic coherence length, since the $T$$\mathrm{_{C}}$ values
we observe are close to the bulk value of $\sim$ 21~K.\cite
{Verdaguer3b, Sakata1} It is important to emphasize that the
primordial ferrimagnetic domains possess net magnetic moments
which have been aligned parallel to H$_{\mathrm{E}}$.
Consequently, the net magnetic field acting on the diamagnetic
sites is the vector sum of the external magnetic field
(H$\mathrm{_{E}}$) and the dipolar field (H$\mathrm{_{D}}$), where
H$\mathrm{_{D}}$ is antiparallel to H$\mathrm{_{E}}$.  When
illuminated, the diamagnetic site will become ferrimagnetic, and
the direction of the photoinduced magnetization will follow that
of the net field as shown in Fig.~2(b) (where the H$_{\mathrm{D}}$
$>$ H$_{\mathrm{E}}$), and Fig.~2(c) (where the H$_{\mathrm{D}}$
$<$ H$_{\mathrm{E}}$).  As a result, the net magnetization of the
film in the photoinduced state \emph{decreases} when
H$_{\mathrm{D}}$ $>$ H$_{\mathrm{E}}$, and this expectation is
consistent with our result for the photoinduced decrease of the
magnetization as shown in Figs.~1(c) and 1(d). When
H$_{\mathrm{D}}$ $<$ H$_{\mathrm{E}}$ as shown in Fig.~2(c), the
net magnetization is expected to increase, and this effect was
observed experimentally at H$_{\mathrm{E}}$ = 7~T, see Fig.~2(d).

For H$_{\mathrm{E}}$ $\parallel$ film, the net moments of the
primordial ferrimagnetic domains align with H$_{\mathrm{E}}$.
Consequently, due to the two-dimensional nature of the film, the
diamagnetic sites experience a net magnetic field which is always
in the direction of H$_{\mathrm{E}}$, and the magneto-optically
induced diamagnetic to ferrimagnetic conversion results in a
global increase in the total magnetization, Figs.~1(a) and 1(b).

In summary, using a sequential assembly method, we synthesized a
magneto-optically active thin film of
Rb$_{j}$Co$_{k}$[Fe(CN)$_{6}$]$_{l}$$\cdot$$n$H$_{2}$O.   Below
$T$$_{\mathrm{C}}$ $\sim$ 20~K when the film was placed parallel
to the external magnetic field, the net magnetization increased
upon irradiation with light.  However, when the film was oriented
perpendicular to the external field, the net magnetization in the
photoinduced state \emph{decreased}. The anisotropy of these
magneto-optical properties arises from an interplay between the
low-dimensional nature of the system and the dipolar fields. The
ability to increase or decrease the photoinduced magnetization by
changing the orientation of the system with respect to the field
is a new phenomenon, heretofore not observed in magnetic thin
films, and provides a novel mechanism
that may be useful in future device applications.\\ \\
%\begin{acknowledgments}

This work was supported, in part, by NSF DMR-9900855 (DRT), NSF
DMR-0113714 (MWM and DRT), ACS-PRF 36163-AC5 (MWM and DRT), NSF
DGE-0209410 (EC), and NSF DMR-0305371 (MWM).  We gratefully
acknowledge early contribution from J. T. Culp.
%\end{acknowledgments}

\begin{thebibliography}{99}

\bibitem[a)]{Park}Electronic mail: juhyun@phys.ufl.edu

\bibitem[b)]{Cizmar}Current address: Institute of Physics, Faculty
of Science, P. J. \v{S}af\'{a}rik University, Ko\v{s}ice,
Slovakia.

\bibitem[c)]{Huh}Current address: Department of Chemistry,
Dankook University, Seoul, Korea.

%1
\bibitem{Verdaguer1}A. Goujon, O. Roubeau, F. Varret, A.~Dolbecq, A.~Bleuzen,
and M. Verdaguer, Eur. Phys. J. B \textbf{14}, 115 (2000).

%2
\bibitem{Villain1}M. Verdaguer, A. Bleuzen, V. Marvaud, J. Vaissermann,
M. Seuleiman, C. Desplanches, A.~Scuiller, C. Train, R. Garde, G.
Gelly, C. Lomenech, I. Rosenman, P. Veillet, C. Cartier, and F.~
Villain, Coord. Chem. Rev. \textbf{190-192}, 1023 (1999).

%3
\bibitem{Talham1}J. T. Culp, J.-H. Park, I. O. Benitez, Y. D. Huh, M. W. Meisel,
and D. R. Talham, Chem.~Mater. \textbf{15}, 3431 (2003).
%4
\bibitem{Talham2}J. T. Culp, J.-H. Park, I. O. Benitez, M. W. Meisel, and D. R. Talham, Polyhedron
\textbf{22}, 2125 (2003).

%5
\bibitem{Verdaguer3a}A. Bleuzen, C. Lomenech, V. Escax, F. Villain,
F. Varret, C. Cartier dit Moulin, and \mbox{M.~Verdaguer}, J. Am.
Chem. Soc. \textbf{122}, 6648 (2000).

%6
\bibitem{Verdaguer3c}V. Escax, A. Bleuzen, C. Cartier dit Moulin, F. Villain,
A. Goujon, F. Varret, and M. Verdaguer, J. Am. Chem. Soc.
\textbf{123}, 12536 (2001).



%7
\bibitem{Verdaguer2}M. Verdaguer, Science \textbf{272}, 698 (1996).

%8
\bibitem{Hashimoto1}O. Sato, T. Iyoda, A. Fujishima, and K. Hashimoto,
 Science \textbf{272}, 704 (1996).

%9
\bibitem{Yamabe1}K. Yoshizawa, F. Mohri, G. Nuspl, and T. Yamabe,
J. Phys. Chem. B \textbf{102}, 5432 (1998).

%10
\bibitem{Hashimoto2}O. Sato, Y. Einaga, A. Fujishima,
 and K. Hashimoto, Inorg. Chem. \textbf{38}, 4405 (1999).

%11
\bibitem{Epstein1}D. A. Pejakovi\'{c} , J. L. Manson, J. S. Miller, and A. J. Epstein,
Phys. Rev. Lett. \textbf{85}, 1994 (2000).


%12
\bibitem{Verdaguer3b}C. Cartier dit Moulin, F.~Villain, A.
Bleuzen, M.-A. Arrio, P.~Sainctavit, C. Lomenech, V.~Escax, F.~
Baudelet, E. Dartyge, J. J. Gallet, and M.~Verdaguer, J.~Am. Chem.
Soc. \textbf{122}, 6653 (2000).

%13
\bibitem{Verdaguer3d}G. Champion, V. Escax, C. Cartier
dit Moulin, A. Bleuzen, F. Villain, F. Baudelet, E. Dartyge, and
M. Verdaguer, J. Am. Chem. Soc. \textbf{123}, 12544 (2001).

%14
\bibitem{Abe1}T. Kawamoto, Y. Asai, and S. Abe, Phys. Rev. Lett.
\textbf{86}, 348 (2001).

%15
\bibitem{Miller1}F. Varret, M. Nogues, and A. Goujon in \emph{Magnetism:
Molecules to Materials I}, edited by J.~S.~Miller and M.~Drillon
(WILEY-VCH Verlag GmbH, Weinheim, 2001), p. 257.

%16
\bibitem{Hashimoto4}S. Ohkoshi and K. Hashimoto, J. Photochem. Photobiol. C
\textbf{2}, 71 (2001).

%17
\bibitem{Hashimoto3}H. Tokoro, S. Ohkoshi, and K. Hashimoto, Appl.
Phys. Lett. \textbf{82}, 1245 (2003).

%18
\bibitem{Sakata1}M. Hanawa, Y. Moritomo, A. Kuriki, J. Tateishi, K. Kato, M. Takata, and M. Sakata,
J. Phys. Soc. Jpn. \textbf{72}, 987 (2003).

%19
\bibitem{Park1}J.-H. Park, Y. D. Huh, E. \v{C}i\v{z}m\'{a}r, S. J. Gamble,
\mbox{D. R.~Talham}, and M. W. Meisel. J. Magn. Magn. Mater.
\textbf{272-276}, 1116 (2004).


%20
\bibitem{FieldCool}The choice of field-cooling instead of
zero-field-cooling was made to maximize the magnetization along
the external field and to minimize the variation of the
magnetization due to the irreversible temperature dependence of
the zero-field-cooled magnetization, an effect which is a common
characteristic of a spin-glass.




\end{thebibliography}

\newpage
%\begin{reference}

\newpage
\begin{table}
\caption{\label{tab:table1}Spin configurations of
Rb$_{j}$Co$_{k}$[Fe(CN)$_{6}$]$_{l}$$\cdot$$n$H$_{2}$O when $T$
$<$ $T$$\mathrm{_{C}}$.}
\begin{ruledtabular}
\begin{tabular}{lcr}
States &Spin Configurations \footnote{LS $\equiv$ low spin and HS
$\equiv$ high spin.}\\ \hline \\ [-3mm]
Diamagnetic & Fe$\mathrm{^{II}}$(LS, $S = 0$)--CN--Co$\mathrm{^{III}}$(LS, $S = 0$) \\
Ferrimagnetic & Fe$\mathrm{^{III}}$(LS, $S = 1/2$)--CN--Co$\mathrm{^{II}}$(HS, $S = 3/2$) \\
\end{tabular}
\end{ruledtabular}
\end{table}

\clearpage
\newpage
\begin{figure}[h]
\caption{Time and temperature dependences of the magnetization of
photoinduced (light) and dark states, when the external magnetic
field is parallel to the film (a, b) and perpendicular to the film
(c, d).}
\end{figure}

\begin{figure}[h]
\caption{Schematic description of the spin configurations of the
domains in the film when \mbox{H$_{\mathrm{E}}$ $\perp$ surface},
where H$_{\mathrm{E}}$ is in the $+$$\hat{z}$ direction. Here, the
clear and the shaded arrows represent Co$\mathrm{^{II}}$(HS, $S =
3/2$) and Fe$\mathrm{^{III}}$(LS, $S = 1/2$) respectively. In (a),
the primordial ferrimagnetic domains coexist with diamagnetic
regions when cooled to $T$ $<$ $T$$_{\mathrm{C}}$ before
irradiation. The net magnetic field on the diamagnetic site is a
vector sum of the external magnetic field (H$_{\mathrm{E}}$) and
the dipolar field (H$_{\mathrm{D}}$) produced by the ferrimagnetic
domains. In (b), at $T$ $<$ $T$$_{\mathrm{C}}$ and
H$_{\mathrm{D}}$ $>$ H$_{\mathrm{E}}$, the diamagnetic regions are
magneto-optically converted to ferrimagnetic ones, and the newly
formed ferrimagnetic net moments point in the $-$$\hat{z}$
direction, resulting in a \emph{decrease} of the net
magnetization. In (c), the photoinduced ferrimagnetic net moments
point $+$$\hat{z}$ direction since H$_{\mathrm{D}}$ $<$
H$_{\mathrm{E}}$, and in this case, the net magnetization
increases in the photoinduced state, as observed experimentally
when H$_{\mathrm{E}}$ = 7~T, see (d).}
\end{figure}

\clearpage
\newpage
\begin{figure}[h]
\includegraphics[width=2.5in,keepaspectratio]{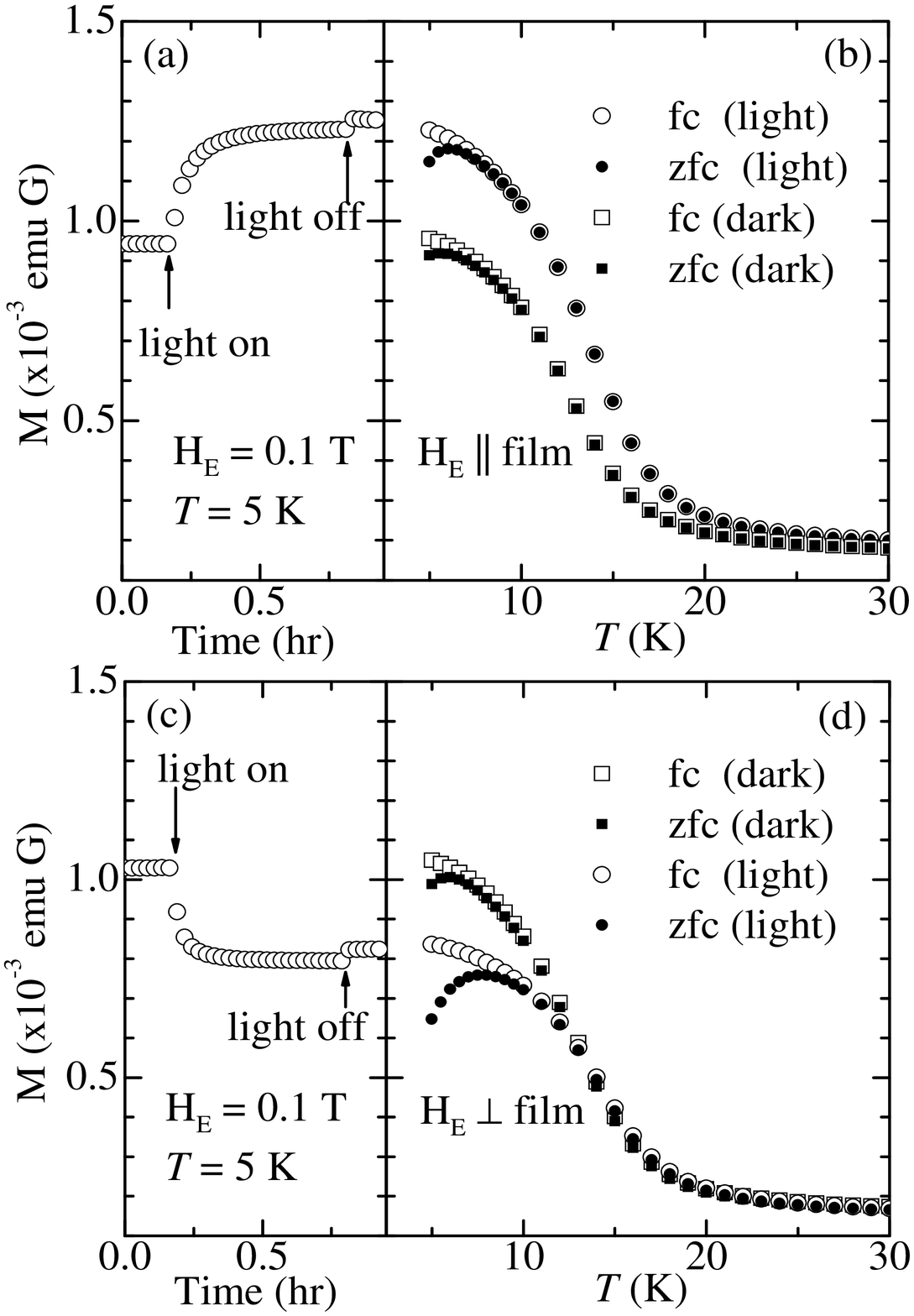}
\end{figure}
\vspace*{0.5cm} \normalsize \centering
Figure 1. J.-H. Park,
Applied Physics Letters.

\newpage
\begin{figure}[h]
\includegraphics[width=2.5in,keepaspectratio,clip]{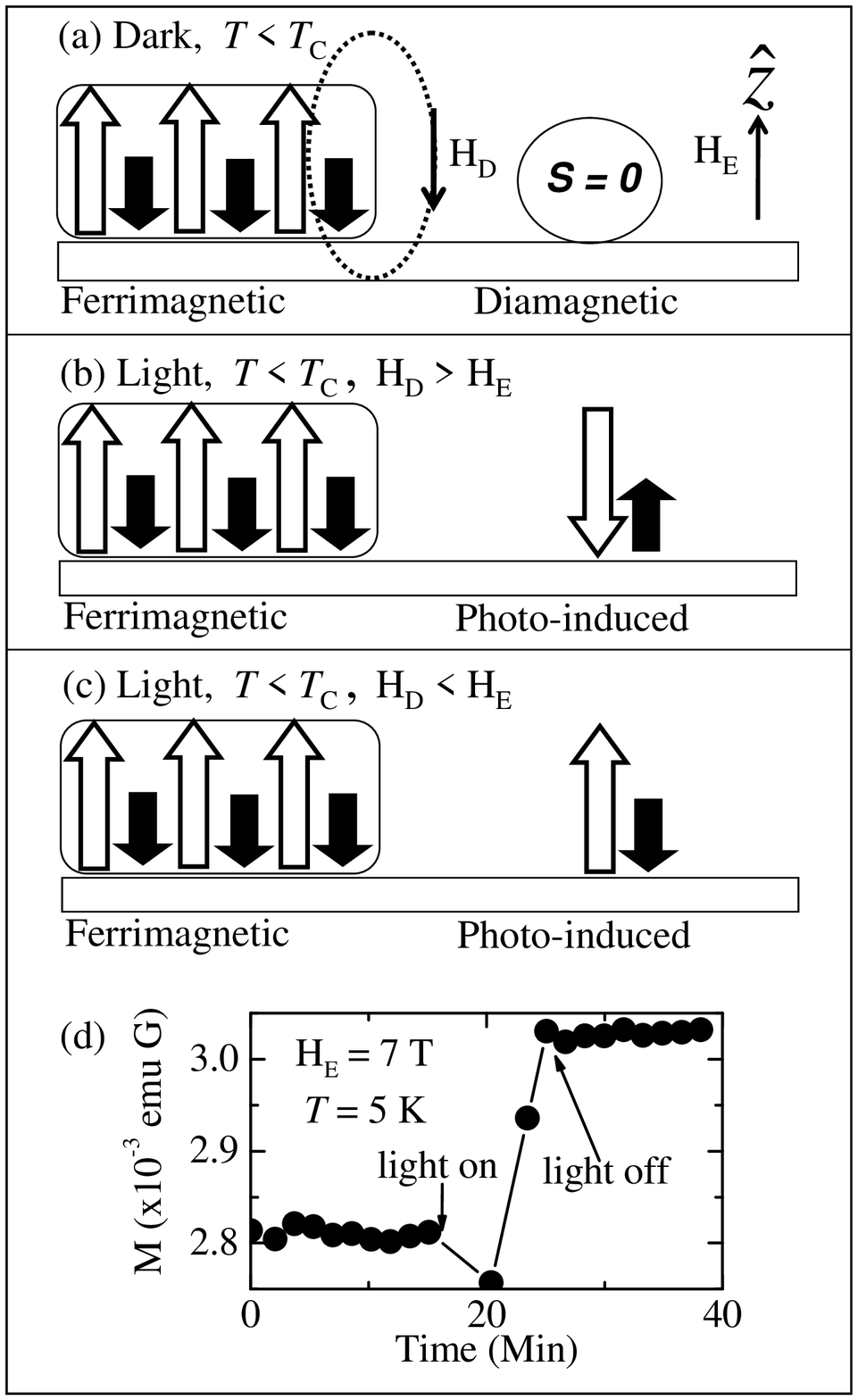}
\end{figure}
\vspace*{0.5cm} \normalsize \centering
Figure 2. J.-H. Park,
Applied Physics Letters.

\end{document}